\documentclass[sigconf]{acmart}

\usepackage{booktabs} % For formal tables
\usepackage{csquotes}
\usepackage{balance}

\usepackage{framed}
\usepackage{subfig}
\usepackage{amsmath}
\usepackage{amsthm}
\usepackage{amssymb}
\usepackage{tabularx}
\usepackage[export]{adjustbox}
\usepackage{rotating}
\usepackage{multirow}
\usepackage{xspace}
\usepackage{tabularx}
\usepackage{enumitem}
\setlist[description]{leftmargin=0.2cm,labelindent=0cm}
\usepackage{pgf}

\newcommand{\palpha}{\ensuremath{\text{P}^3\!\alpha}}
\newcommand{\pbeta}{\ensuremath{\text{RP}^3\!\beta}}
\renewcommand{\vec}[1]{\mathbf{#1}}

\AtBeginDocument{%
  \providecommand\BibTeX{{%
    \normalfont B\kern-0.5em{\scshape i\kern-0.25em b}\kern-0.8em\TeX}}}

\setcopyright{acmcopyright}

\copyrightyear{2019}
\acmYear{2019}
\acmConference[RecSys '19]{Thirteenth ACM Conference on Recommender Systems}{September 16--20, 2019}{Copenhagen, Denmark}
\acmBooktitle{Thirteenth ACM Conference on Recommender Systems (RecSys '19), September 16--20, 2019, Copenhagen, Denmark}
\acmPrice{15.00}
\acmDOI{10.1145/3298689.3347058}
\acmISBN{978-1-4503-6243-6/19/09}

\usepackage{booktabs} % For formal tables

\begin{document}

\title[Are We Really Making Much Progress?]{Are We Really Making Much Progress? A Worrying Analysis of Recent Neural Recommendation Approaches}

\author{Maurizio Ferrari Dacrema}
\orcid{0000-0001-7103-2788}
\affiliation{%
  \institution{Politecnico di Milano, Italy}
}
\email{maurizio.ferrari@polimi.it}

\author{Paolo Cremonesi}
\orcid{0000-0002-1253-8081}
\affiliation{%
  \institution{Politecnico di Milano, Italy}
}
\email{paolo.cremonesi@polimi.it}

\author{Dietmar Jannach}
\orcid{0000-0002-4698-8507}
\affiliation{%
  \institution{University of Klagenfurt, Austria}
}
\email{dietmar.jannach@aau.at}

\begin{abstract}
Deep learning techniques have become the method of choice for researchers working on algorithmic aspects of recommender systems. With the strongly increased interest in machine learning in general, it has, as a result, become difficult to keep track of what represents the state-of-the-art at the moment, e.g., for top-n recommendation tasks.
At the same time, several recent publications point out problems in today's research practice in applied machine learning, e.g., in terms of the reproducibility of the results or the choice of the baselines when proposing new models.

In this work, we report the results of a systematic analysis of algorithmic proposals for top-n recommendation tasks.  
Specifically, we considered 18 algorithms that were presented at top-level research conferences in the last years.
Only 7 of them could be reproduced with reasonable effort.
For these methods, it however turned out that 6 of them can often be outperformed with comparably simple heuristic methods, e.g., based on nearest-neighbor or graph-based techniques.
The remaining one clearly outperformed the baselines 
but did not consistently outperform a well-tuned non-neural linear ranking method.
Overall, our work sheds light on a number of potential problems in today's machine learning scholarship and calls for improved scientific practices in this area.
\end{abstract}



\maketitle

\section{Introduction}
Within only a few years, deep learning techniques have started to dominate the landscape of algorithmic research in recommender systems. Novel methods were  
proposed for a variety of settings and algorithmic tasks, including top-n recommendation based on long-term preference profiles or for session-based recommendation scenarios \cite{QuadranaetalCSUR2018}.
Given the increased interest in machine learning in general, the corresponding number of recent research publications, and the success of deep learning techniques in other fields like vision or language processing, one could expect that substantial progress resulted from these works 
also in the field of recommender systems.
However, indications exist in other application areas of machine learning that the achieved progress---measured in terms of accuracy improvements over existing models---is not always as strong as expected. 

Lin \cite{Lin:2019:NHC:3308774.3308781}, for example, discusses two recent neural approaches in the field of information retrieval that were published at top-level conferences. His analysis reveals that the new methods do \emph{not} significantly outperform existing baseline methods when these are carefully tuned. In the context of recommender systems, an in-depth analysis presented in \cite{Ludewig2018} shows that even a very recent neural method for session-based recommendation can, in most cases, be outperformed by very simple methods based, e.g., on nearest-neighbor techniques. Generally, questions regarding the true progress that is achieved in such applied machine learning settings are not new, nor tied to research based on deep learning. Already in 2009, Armstrong et al.~\cite{Armstrong:2009:IDA:1645953.1646031} concluded from an analysis in the context of ad-hoc retrieval tasks that, despite many papers being published, the reported improvements ``don't add up''.

Different factors contribute to such phenomena, including (i) weak baselines; (ii) establishment of weak methods as new baselines; and (iii) difficulties in comparing or reproducing results across papers. 
One first problem lies in the choice of the baselines that are used in the comparisons.
Sometimes, baselines are chosen that are too weak in general for the given task and dataset, and sometimes the baselines are not properly fine-tuned.
Other times, baselines are chosen from the same family as the newly proposed algorithm, e.g., when a new deep learning algorithm is compared only against other deep learning baselines.
This behaviour enforces the propagation of weak baselines. When previous deep learning algorithms were evaluated against too weak baselines, the new deep learning algorithm will not necessarily improve over strong non-neural baselines. Furthermore, with the constant flow of papers being published in recent years, keeping track of what represents a state-of-the-art baseline becomes increasingly challenging.

Besides issues related to the baselines, an additional challenge is that researchers use various types of datasets, evaluation protocols, performance measures, and data preprocessing steps, which makes it difficult to conclude which method is the best across different application scenarios. This is in particular problematic when source code and data are not shared. 
While we observe an increasing trend that researchers publish the source code of their algorithms, this is not the common rule today even for top-level publication outlets. And even in cases when the code is published, it is sometimes incomplete and, for instance, does not include the code for data preprocessing, parameter tuning, or the exact evaluation procedures, as pointed out also in \cite{DBLP:conf/aaai/0002IBPPM18}.

Finally, another general problem might lie in today's research practice in applied machine learning in general. Several ``troubling trends'' are discussed in \cite{troubling-trends-1807.03341}, including the thinness of reviewer pools or misaligned incentives for authors that might stimulate certain types of research. Earlier work \cite{DBLP:journals/corr/abs-1206-4656} also discusses the community's focus on abstract accuracy measures or the narrow focus of machine learning research in terms of what is ``publishable'' at top publication outlets.

With this research work, our goal is to shed light on the question if the problems reported above also exist in the domain of deep learning-based recommendation algorithms. Specifically, we address two main research questions:
\begin{enumerate}
\item \emph{Reproducibility}: To what extent is recent research in the area reproducible (with reasonable effort)?
\item \emph{Progress}: To what extent are recent algorithms actually leading to better performance results when compared to relatively simple, but well-tuned, baseline methods?
\end{enumerate}
To answer these questions, we conducted a systematic study in which we analyzed research papers that proposed new algorithmic approaches for top-n recommendation tasks using deep learning methods. To that purpose, we scanned the recent conference proceedings of KDD, SIGIR, TheWebConf (WWW), and RecSys for corresponding research works. We identified 18 relevant papers.

In a first step, we tried to reproduce the results reported in the paper for those cases where the source code was made available by the authors and where we had access to the data used in the experiments. In the end, we could reproduce the published results with an acceptable degree of certainty for only 7 papers. A first contribution of our work is therefore an assessment of the reproducibility level of current research in the area. 

In the second part of our study, we re-executed the experiments reported in the original papers, but also included additional baseline methods in the comparison. Specifically, we used heuristic methods based on user-based and item-based nearest neighbors as well as two variants of a simple graph-based approach. 
Our study, to some surprise, revealed that in the large majority of the investigated cases (6 out of 7) the proposed deep learning techniques did not consistently outperform the simple, but fine-tuned, baseline methods. In one case, even a non-personalized method that recommends the most popular items to everyone was the best one in terms of certain accuracy measures. Our second contribution therefore lies in the identification of a potentially more far-reaching problem related to current research practices in machine learning.

The paper is organized as follows. Next, in Section \ref{sec:research-method}, we describe our research method and how we reproduced existing works.
The results of re-executing the experiments while including additional baselines are provided in Section \ref{sec:comparison}. We finally discuss the implications of our research in Section \ref{sec:discussion}.

\section{Research Method}
\label{sec:research-method}
\subsection{Collecting Reproducible Papers}
To make sure that our work is not only based on individual examples of recently published research, we systematically scanned the proceedings of scientific conferences for relevant long papers in a manual process.
Specifically, we included long papers in our analysis that appeared between 2015 and 2018 in the following four conference series: KDD, SIGIR, TheWebConf (WWW), and RecSys.\footnote{All of the conferences are either considered A* in the Australian Core Ranking or specifically dedicated to research in recommender systems.} We considered a paper to be relevant if it (a) proposed a deep learning based technique and (b) focused on the top-n recommendation problem. Papers on other recommendation tasks, e.g., group recommendation or session-based recommendation, were not considered in our analysis. 
Given our interest in top-n recommendation, we considered only papers that used for evaluation classification or ranking metrics, such as Precision, Recall, MAP.
After this screening process, we ended up with a collection of 18 relevant papers.

In a next step, we tried to reproduce\footnote{Precisely speaking, we used a mix of replication and reproduction \cite{Plesser2018,ACMreproducibilty2016}, i.e., we used both artifacts provided by the authors and our own artifacts. For the sake of readability, we will only use the term ``reproducibility'' in this paper.} the results reported in these papers. Our approach to reproducibility is to rely as much as possible on the artifacts provided by the authors themselves, i.e., their source code and the data used in the experiments. In theory, it should be possible to reproduce published results using only the technical descriptions in the papers. In reality, there are, however many tiny details regarding the implementation of the algorithms and the evaluation procedure, e.g., regarding data splitting, that can have an impact on the experiment outcomes \cite{Said:2014:RTF:2645710.2645712}.

We therefore tried to obtain the code and the data for all relevant papers from the authors. In case these artifacts were not already publicly provided, we contacted all authors of the papers and waited 30 days for a response. In the end, we considered a paper to be \emph{reproducible}, if the following conditions were met:
\begin{itemize}
    \item A working version of the \emph{source code} is available or the code only has to be modified in minimal ways to work correctly.\footnote{We did not apply modifications to the core algorithms.}
    \item At least one \emph{dataset} used in the original paper is available. A further requirement here is that either the originally-used train-test splits are publicly available or that they can be reconstructed based on the information in the paper.
\end{itemize}

Otherwise, we consider a paper to be \emph{non-reproducible} given our specific reproduction approach. Note that we also considered works to be non-reproducible when the source code was published but contained only a skeleton version of the model with many parts and details missing. Concerning the datasets, research based solely on non-public data owned by companies or data that was gathered in some form from the web but not shared publicly, was also not considered reproducible.

The fraction of papers that were reproducible according to our relatively strict criteria per conference series
are shown in Table \ref{tab:reproducibility-stats}.

\begin{table}[h!t]
\centering
\caption{Reproducible works on deep learning algorithms for top-n recommendation per conference series from 2015 to 2018.}
\label{tab:reproducibility-stats}
\begin{tabular}{lll}
  Conference  & Rep. ratio  & Reproducible \\ \midrule
  KDD  & 3/4 (75\%) &  \cite{hu2018leveragingmetapathcontext}, \cite{li2017collaborativevariationalautoencoder}, \cite{wang2015collaborativedeeplearning} \\
  RecSys  & 1/7 (14\%)& \cite{Zheng:2018:SCF:3240323.3240343} \\
  SIGIR & 1/3 (30\%)& \cite{ebesu2018collaborative} \\
  WWW  & 2/4 (50\%)& \cite{he2017neural}, \cite{liang2018variationalautoencodersforCF} \\ \midrule
  Total & 7/18 (39\%) & \\
  \bottomrule
  \multicolumn{3}{p{0.8\columnwidth}}{\emph{Non-reproducible:} KDD: \cite{tay2018multipointercoattention}, RecSys: \cite{Sun:2018:RKG:3240323.3240361}, \cite{Bharadhwaj:2018:RRG:3240323.3240383}, \cite{Sachdeva:2018:ANA:3240323.3240397}, \cite{Tuan:2017:CNS:3109859.3109900}, \cite{Kim:2016:CMF:2959100.2959165},  \cite{Vasile:2016:MPE:2959100.2959160}, SIGIR: \cite{manotumruksa2018contextualattention}, \cite{chen2017attentivecomponentlevelattention}, WWW: \cite{tay2018latentrelationalmetric}, \cite{elkahky2015multiviewdeeplearningcrossdomain}}  \\
\end{tabular}
\end{table}

Overall, we could reproduce only about one third of the works, which confirms previous discussions about limited reproducibility, see, e.g., \cite{Beel2016}.
The sample size is too small to make reliable conclusions regarding the difference between conference series. The detailed statistics per year---not shown here for space reasons---however indicate that the reproducibility rate increased over the years.

\subsection{Evaluation Methodology}

\paragraph{Measurement Method}
The validation of the progress that is achieved through new methods against a set of baselines can be done in at least two ways. One is to evaluate all considered methods within the same  defined environment, using the same datasets and the exact same evaluation procedure for all algorithms as done in \cite{Ludewig2018}. While such an approach helps us obtain a picture of how different methods compare across datasets, the implemented evaluation procedure might be slightly different from the one used in the original papers. As such, this approach would not allow us to \emph{exactly} reproduce what has been originally reported, which is the goal in this present work.

In this work, we therefore reproduce the work by refactoring the original implementations in a way that allows us to apply the same evaluation procedure that was used in the original papers. Specifically, refactoring is done in a way that the original code for training, hyper-parameter optimization and prediction are separated from the evaluation code. This evaluation code is then also used for the baselines.

For all reproduced algorithms considered in the individual experiments, we used the optimal hyper-parameters that were reported by the authors in the original papers for each dataset. This is appropriate because we used the same datasets, algorithm implementation, and evaluation procedure as in the original papers.\footnote{We will re-run parameter optimization for the reproduced algorithms as part of our future work in order to validate the parameter optimization procedures used by the authors. This step was, however, outside the scope of our current work.} We share all the code and data used in our experiments as well as details of the final algorithm (hyper-)parameters of our baselines along with the full experiment results online. \footnote{\url{https://github.com/MaurizioFD/RecSys2019_DeepLearning_Evaluation}}

\paragraph{Baselines}
We considered the following baseline methods in our experiments, all of which are conceptually simple.
\begin{description}
\item[TopPopular:]
A non-personalized method that recommends the most popular items to everyone. Popularity is measured by the number of explicit or implicit ratings.

\item[ItemKNN:]
A traditional Collaborative-Filtering (CF) approach based on $k$-nearest-neighborhood (KNN) and item-item similarities \cite{wang2006unifying}.
We used the cosine similarity $s_{ij}$ between items $i$ and $j$ computed as
\begin{equation}
\label{eqn:cosine_similarity}
s_{ij} = \frac{\vec{r}_i \cdot \vec{r}_j}{\| \vec{r}_i\| \| \vec{r}_j\| + h}
\end{equation}
where vectors $\vec{r}_i, \vec{r}_j \in \mathbb{R} ^ {|U|}$ represent the implicit ratings of a user for items $i$ and $j$, respectively, and
$|U|$ is the number of users.
Ratings can be optionally weighted either with TF-IDF or BM25, as described in \cite{wang2008probabilistic}. Furthermore the similarity may or not be normalized via the product of vector norms.
Parameter $h$ (the \emph{shrink term}) is used to lower the similarity between items having only few interactions \cite{bell2007improved}.
The other parameter of the method is the neighborhood size $k$.

\item[UserKNN:]
A neighborhood-based method using collaborative user-user similarities.
Hyper-parameters are the same as used for ItemKNN \cite{sarwar2001item}.

\item[ItemKNN-CBF:]
A neighborhood content-based-filtering (CBF) approach with item similarities computed by using item content features (attributes)
\begin{equation}
\label{eqn:cosine_similarity2}
s_{ij} = \frac{\vec{f}_i \cdot \vec{f}_j}{\| \vec{f}_i\| \| \vec{f}_j\| + h}
\end{equation}
where vectors $\vec{f}_i, \vec{f}_j \in \mathbb{R} ^ {|F|}$ describe the features of items $i$ and $j$, respectively, and
$|F|$ is the number of features.
Features can be optionally weighted either with TF-IDF or BM25.
Other parameters are the same used for ItemKNN \cite{lops2011content}.

\item[ItemKNN-CFCBF:]
A hybrid CF+CFB algorithm based on item-item similarities.
The similarity is computed by first concatenating, for each item $i$, the vector of ratings and the vector of features --
$\left[ \vec{r}_i \, , w \vec{f}_i \right]$ -- and by later computing the cosine similarity \label{eqn:cosine_similarity2} between the concatenated vectors.
Hyper-parameters are the same used for ItemKNN, plus a parameter $w$ that weights the content features with respect to the ratings.

\item[\palpha :]
A simple graph-based algorithm which implements a random walk between users and items \cite{cooper2014P3alpha}.
Items for user $u$ are ranked based on the probability of a random walk with three steps starting from user $u$.
The probability $p_{ui}$ to jump from user $u$ to item $i$ is computed from the implicit user-rating-matrix as
$p_{ui} = \left( r_{ui} / N_u \right) ^\alpha $, where $r_{ui}$ is the rating of user $u$ on item $i$, $N_u$ is the number of ratings of user $u$ and $\alpha$ is a damping factor.
The probability $p_{iu}$ to jump backward is computed as $p_{iu} = \left( r_{ui} / N_i \right) ^\alpha $, where $N_i$ is the number of ratings for item $i$.
The method is equivalent to a KNN item-based CF algorithm, with the similarity matrix defined as
\begin{equation}
\label{eqn:p3a}
s_{ij} = \sum_v{ p_{jv} p_{vi} }
\end{equation}

The parameters of the method are the numbers of neighbors $k$ and the value of $\alpha$.
We include this algorithm because it provides good recommendation quality at a low computational cost.

\item[\pbeta :]
A version of \palpha~proposed in \cite{paudel2017Rp3beta}. Here, the outcomes of \palpha~are modified by dividing the similarities by each item's popularity raised to the power of a coefficient $\beta$. If $\beta$ is 0, the algorithm is equivalent to \palpha.
Its parameters are the numbers of neighbors $k$ and the values for $\alpha$ and $\beta$.
\end{description}

For all baseline algorithms and datasets, we determined the optimal parameters via Bayesian search \cite{antenucci2018artist} using the implementation of Scikit-Optimize\footnote{\url{https://scikit-optimize.github.io/}}.
We explored 35 cases for each algorithm, where the first 5 were used for the initial random points.
We considered neighborhood sizes $k$ from 5 to 800; the shrink term $h$ was between 0 and 1000; and $\alpha$ and $\beta$ took real values between 0 and 2.

\section{Validation Against Baselines}
\label{sec:comparison}
This section summarizes the results of comparing the reproducible works with the described baseline methods. We share the detailed statistics, results, and final parameters online.

\subsection{Collaborative Memory Networks (CMN)}
\label{subsec:cmf}
The CMN method was presented at SIGIR '18 and combines memory networks and neural attention mechanisms with latent factor and neighborhood models \cite{ebesu2018collaborative}. To evaluate their approach, the authors compare it with different matrix factorization and neural recommendation approaches as well as with an ItemKNN algorithm (with no \emph{shrinkage}). Three datasets are used for evaluation: Epinions, CiteULike-a, and Pinterest. Optimal hyper-parameters for the proposed method are reported, but no information is provided on how the baselines are tuned. Hit rate and NDCG are the performance measures used in a leave-one-out procedure. The reported results show that CMNs outperform all other baselines on all measures.

We were able to reproduce their experiments for all their datasets. For our additional experiments with the simple baselines, we optimized the parameters of our baselines for the hit rate (HR@5) metric. The results for the three datasets are shown in Table \ref{tab:CMN-results}.

\begin{table}[h!t]
    \caption{Experimental results for the CMN method using the metrics and cutoffs reported in the original paper. Numbers are printed in bold when they correspond to the best result or when a baseline outperformed CMN.}
    \label{tab:CMN-results}
%\resizebox{.8\linewidth}{!}{%
    \begin{tabular}{lcccccc}
   \toprule
	& \multicolumn{4}{c}{CiteULike-a}  	\\
		& HR@5 	& NDCG@5 	& HR@10 	& NDCG@10 	\\
\cmidrule(lr){2-5}
		%\midrule
%    Random	&0.0562	&0.0332	&0.1007	&0.0474	\\
    TopPopular	&0.1803	&0.1220	&0.2783	&0.1535	\\
    UserKNN	&\textbf{0.8213}	&\textbf{0.7033}	&\textbf{0.8935}	&\textbf{0.7268}	\\
    ItemKNN	&\textbf{0.8116}	&\textbf{0.6939}	&0.8878	&\textbf{0.7187}	\\
    \smash{\palpha}	&\textbf{0.8202}	&\textbf{0.7061}	&0.8901	&\textbf{0.7289}	\\
    \smash{\pbeta}	&\textbf{0.8226}	&\textbf{0.7114}	&\textbf{0.8941}	&\textbf{0.7347}	\\
%    PureSVD	&0.7056	&0.5807	&0.7892	&0.6079	\\
    \midrule
    CMN &0.8069	&0.6666	&0.8910	&0.6942	\\
    \midrule
	& \multicolumn{4}{c}{Pinterest}  	\\
		& HR@5 	& NDCG@5 	& HR@10 	& NDCG@10 	\\
		\cmidrule(lr){2-5}
%\midrule
%    Random	&0.0495	&0.0293	&0.0982	&0.0449	\\
    TopPopular	&0.1668	&0.1066	&0.2745	&0.1411	\\
    UserKNN	&\textbf{0.6886}	&\textbf{0.4936}	&0.8527	&\textbf{0.5470}	\\
    ItemKNN	&\textbf{0.6966}	&\textbf{0.4994}	&\textbf{0.8647}	&\textbf{0.5542}	\\
    \smash{\palpha}	&0.6871	&\textbf{0.4935}	&0.8449	&\textbf{0.5450}	\\
    \smash{\pbeta}	&\textbf{0.7018}	&\textbf{0.5041}	&\textbf{0.8644}	&\textbf{0.5571}	\\
%    PureSVD	&0.6440	&0.4565	&0.8064	&0.5095	\\
    \midrule
    CMN	&0.6872	&0.4883	&0.8549	&0.5430	\\
    \midrule
	& \multicolumn{4}{c}{Epinions}  	\\
		& HR@5 	& NDCG@5 	& HR@10 	& NDCG@10 	\\
		%\midrule
\cmidrule(lr){2-5}

%    Random	&0.0476	&0.0277	&0.0965	&0.0433	\\
    TopPopular	&\textbf{0.5429}	&\textbf{0.4153}	&\textbf{0.6644}	&\textbf{0.4547}	\\
    UserKNN	&0.3506	&0.2983	&0.3922	&0.3117	\\
    ItemKNN	&0.3821	&0.3165	&0.4372	&0.3343	\\
    \smash{\palpha}	&0.3510	&0.2989	&0.3891	&0.3112	\\
    \smash{\pbeta}	&0.3511	&0.2980	&0.3892	&0.3103	\\
%    PureSVD	&0.3715	&0.2989	&0.4473	&0.3234	\\
    \midrule
    CMN &0.4195	&0.3346	&0.4953	&0.3592	\\
	\bottomrule
  	\end{tabular}
 % 	}
\end{table}

Our analysis shows that, after optimization of the baselines, CMN\footnote{We report the results for CMN-3 as the version with the best results.} is in no single case the best-performing method on any of the datasets. For the CiteULike-a and Pinterest datasets, at least two of the personalized baseline techniques outperformed the CMN method on any measure. Often, even all personalized baselines were better than CMN. For the Epinions dataset, to some surprise, the unpersonalized TopPopular method, which was not included in the original paper, was better than all other algorithms by a large margin. On this dataset, CMN was indeed much better than our baselines. The success of CMN on this comparably small and very sparse dataset with about 660k observations could therefore be tied to the particularities of the dataset or to a popularity bias of CMN. An analysis reveals that the Epinions dataset has indeed a much more uneven popularity distribution than the other datasets (Gini index of 0.69 vs.~0.37 for CiteULike-a). For this dataset, CMN also recommends in its top-n lists items that are, on average, 8\% to 25\% more popular than the items recommended by our baselines.

\subsection{Metapath based Context for RECommendation (MCRec)}
\label{subsec:mcrec}
MCRec \cite{hu2018leveragingmetapathcontext}, presented at KDD '18, is a meta-path based model that leverages auxiliary information like movie genres for top-n recommendation. From a technical perspective, the authors propose a priority-based sampling technique to select higher-quality path instances and propose a novel co-attention mechanism to improve the representations of meta-path based context, users, and items.

The authors benchmark four variants of their method against a variety of models of different complexity on three small datasets (MovieLens100k, LastFm, and Yelp). The evaluation is done by creating 80/20 random training-test splits and by executing 10 of such evaluation runs. The evaluation procedure could be reproduced; public training-test splits were provided only for the MovieLens dataset. For the \emph{MF} and \emph{NeuMF} \cite{he2017neural} baselines used in their paper, the architecture and hyper-parameters were taken from the original papers; no information about hyper-parameter tuning is provided for the other baselines. Precision, Recall, and the NDCG are used as performance measures, with a recommendation list of length 10. The NDCG measure is however implemented in an uncommon and questionable way, which is not mentioned in the paper. Here, we therefore use a standard version of the NDCG.

In the publicly shared software, the meta-paths are hard-coded for MovieLens, and no code for preprocessing and constructing the meta-paths is provided. Here, we therefore only provide the results for the MovieLens dataset in detail. We optimized our baselines for Precision, as was apparently done in \cite{hu2018leveragingmetapathcontext}. For MCRec, the results for the complete model are reported.

\begin{table}[h!t]
    \caption{Comparing MCRec against our baselines (MovieLens100k)}
    \label{tab:MCRec_results}
    \begin{tabular}{lcccccc}
    \toprule
		& PREC@10 	& REC@10 	& NDCG@10 	\\
	\midrule
%    Random	&0.0141	&0.0063	&0.0077	\\
    TopPopular	&0.1907	&0.1180	&0.1361	\\
    UserKNN	&0.2913	&0.1802	&0.2055	\\
    ItemKNN	&\textbf{0.3327}	&\textbf{0.2199}	&\textbf{0.2603}	\\
    \smash{\palpha}	&0.2137	&0.1585	&0.1838	\\
    \smash{\pbeta}	&0.2357	&0.1684	&0.1923	\\
%    PureSVD	&0.1289	&0.1087	&0.1177	\\
    \midrule
%    ItemKNN CBF	&0.0294	&0.0089	&0.0109	\\
%    ItemKNN CFCBF	&\textbf{0.3363}	&\textbf{0.2103}	&\textbf{0.2466}	\\
    MCRec	&0.3077	&0.2061	&0.2363	\\
 	\bottomrule
  	\end{tabular}
 % 	}
\end{table}

Table \ref{tab:MCRec_results} shows that the traditional ItemKNN method, when configured correctly, outperforms MCRec on all performance measures.

Besides the use of an uncommon NDCG measure, we found other potential methodological issues in this paper. Hyper-parameters for the \emph{MF} and \emph{NeuMF} baselines were, as mentioned, not optimized for the given datasets but taken from the original paper \cite{hu2018leveragingmetapathcontext}.
In addition, looking at the provided source code, it can be seen that the authors report the best results of their method for each metric across different epochs chosen on the test set, which is inappropriate.\footnote{In our evaluations, we did not use this form of measurement.}

\subsection{Collaborative Variational Autoencoder (CVAE)}
The CVAE method \cite{li2017collaborativevariationalautoencoder}, presented at KDD '18, is a hybrid technique that considers both content as well as rating information. The model learns deep latent representations from content data in an unsupervised manner and also learns implicit relationships between items and users from both content and ratings.

The method is evaluated on two comparably small CiteULike datasets (135k and 205k interactions).
For both datasets, a sparse and a dense version is tested. The baselines in \cite{li2017collaborativevariationalautoencoder} include three recent deep learning models and as well as Collaborative Topic Regression (CTR).
The parameters for each method are tuned based on a validation set.
Recall at different list lengths (50 to 300) is used as an evaluation measure. Random train-test data splitting is applied and the measurements are repeated five times.

\begin{table}[h!t]
    \caption{Experimental results for CVAE (CiteULike-a).}
    \label{tab:CVAE_results}
    \begin{tabular}{lccc}
    \toprule
		& REC@50 	& REC@100 	& REC@300 	\\
		\midrule
%    Random	&0.0026	&0.0054		&0.0167	\\
    TopPopular	&0.0044	&0.0081		&0.0258	\\
    UserKNN	&0.0683	&0.1016		&0.1685	\\
    ItemKNN	&\textbf{0.0788}	&0.1153	&0.1823	\\
    \smash{\palpha}	&\textbf{0.0788}	&0.1151	&0.1784	\\
    \smash{\pbeta}	&\textbf{0.0811}	&0.1184	&0.1799	\\
%    PureSVD	&0.0515	&0.0771		&0.1380	\\
%    ItemKNN CBF	&\textbf{0.1687}	&\textbf{0.2435}	&\textbf{0.3878}	\\
    ItemKNN-CFCBF	&\textbf{0.1837}	&\textbf{0.2777}&\textbf{0.4486}	\\
    \midrule
    CVAE	&0.0772	&0.1548	&0.3602	\\
	\bottomrule
  	\end{tabular}
\end{table}

We could reproduce their results using their code and evaluation procedure. The datasets are also shared by the authors. Fine-tuning our baselines led to the results shown in Table \ref{tab:CVAE_results} for the dense CiteULike-a dataset from \cite{wang2011collaborativetopicmodeling}.
For the shortest list length of 50, even the majority of the pure CF baselines outperformed the CVAE method on this dataset. At longer list lengths, the hybrid \emph{ItemKNN-CFCBF} method led to the best results. Similar results were obtained for the sparse \emph{CiteULike-t} dataset.  Generally, at list length 50, \emph{ItemKNN-CFCBF} was consistently outperforming CVAE in all tested configurations. Only at longer list lengths (100 and beyond), CVAE was able to outperform our methods on two datasets.

Overall, CVAE was only favorable over the baselines in certain configurations and at comparably long and rather uncommon recommendation cutoff thresholds. The use of such long list sizes was however not justified in the paper.

\subsection{Collaborative Deep Learning (CDL)}
The discussed CVAE method considers the earlier and often-cited CDL method \cite{wang2015collaborativedeeplearning} from KDD '15 as one of their baselines, and the authors also use the same evaluation procedure and CiteULike datasets. CDL is a probabilistic feed-forward model for joint learning of stacked denoising autoencoders (SDAE) and
collaborative filtering. It applies deep learning techniques to jointly learn a deep representation of content information and collaborative information.
The evaluation of CDL in \cite{wang2015collaborativedeeplearning} showed that it is favorable in particular compared to the widely referenced CTR method \cite{wang2011collaborativetopicmodeling}, especially in sparse data situations.

\begin{table}[h!t]
    \caption{Experimental results for CDL on the dense \emph{CiteULike-a} dataset.}
    \label{tab:CDL_results}
    \begin{tabular}{lccc}
    \toprule
		& REC@50 	& REC@100 	& REC@300 	\\
		\midrule
%    Random	&0.0028	&0.0058	&0.0182	\\
    TopPopular	&0.0038	&0.0073	& 0.0258	\\
    UserKNN	&\textbf{0.0685}	&0.1028	&0.1710	\\
    ItemKNN	&\textbf{0.0846}	&\textbf{0.1213}	&0.1861	\\
    \smash{\palpha}	&\textbf{0.0718}	&\textbf{0.1079}		&0.1777	\\
    \smash{\pbeta}	&\textbf{0.0800}	&\textbf{0.1167}	&0.1815	\\
%    PureSVD	&0.0510	&0.0774	& 0.1369	\\
    ItemKNN-CBF	&\textbf{0.2135}	&\textbf{0.3038}		&\textbf{0.4707}	\\
    ItemKNN-CFCBF 	&\textbf{0.1945}	&\textbf{0.2896} &\textbf{0.4620}	\\
    \midrule
    CDL &0.0543	&0.1035	&0.2627	\\
	\bottomrule
  	\end{tabular}
\end{table}

We reproduced the research in \cite{wang2015collaborativedeeplearning}, leading to the results shown in Table \ref{tab:CDL_results} for the dense \emph{CiteULike-a} dataset. Not surprisingly, the baselines that were better than CVAE in the previous section are also better than CDL, and again for short list lengths, already the pure CF methods were better than the hybrid CDL approach. Again, however, CDL leads to higher Recall for list lengths beyond 100 in two out of four dataset configurations. Comparing the detailed results for CVAE and CDL, we see that the newer CVAE method is indeed always better than CDL, which indicates that progress was made. Both methods, however, are not better than one of the simple baselines in the majority of the cases.

\subsection{Neural Collaborative Filtering (NCF)}
Neural network-based Collaborative Filtering \cite{he2017neural}, presented at WWW '17, generalizes Matrix Factorization by replacing the inner product with a neural architecture that can learn an arbitrary function from the data. The proposed hybrid method (NeuMF) was evaluated on two datasets (MovieLens1M and Pinterest), containing 1 million and 1.5 million interactions, respectively. A leave-one out procedure is used in the evaluation and the original data splits are publicly shared by the authors. Their results show that NeuMF is favorable, e.g., over existing matrix factorization models, when using the hit rate
and the NDCG as an evaluation measure using different list lengths up to 10.

Parameter optimization is done on a validation set created from the training set. Similar to the implementation of \emph{MCRec} above, 
the provided source code shows that the authors chose the number of epochs based on the results obtained for the test set.
Since the number of epochs, however, is a parameter to tune and should not be determined based on the test set, we use a more appropriate implementation that finds this parameter with the validation set. For the ItemKNN method, the authors only varied the neighborhood sizes but did not test other variations.

\begin{table}[h!t]
    \caption{Experimental results for NCF.}
    \label{tab:NCF_results}
%\resizebox{\linewidth}{!}{%
    \begin{tabular}{lcccc}
    \toprule
	& \multicolumn{4}{c}{Pinterest}  	\\
	& HR@5 	& NDCG@5 &	 HR@10 	 &NDCG@10 	\\
	%\midrule
\cmidrule(lr){2-5}

%    Random	&0.0100	&0.0100	&0.0490	&0.0290	&0.0978	&0.0446	\\
    TopPopular	&0.1663	&0.1065	&0.2744	&0.1412	\\
    UserKNN	&0.7001	&\textbf{0.5033}	&0.8610	&\textbf{0.5557}	\\
    ItemKNN	&\textbf{0.7100}	&\textbf{0.5092}	&\textbf{0.8744}	&\textbf{0.5629}	\\
    \smash{\palpha}	    &0.7008	&\textbf{0.5018}	&0.8667	&\textbf{0.5559}	\\
    \smash{\pbeta}	 	&\textbf{0.7105}	&\textbf{0.5116}	&\textbf{0.8740}	&\textbf{0.5650}	\\
    \midrule
    NeuMF	    &0.7024	& 0.4983	&0.8719	&0.5536	\\
    \midrule
   & \multicolumn{4}{c}{Movielens 1M} 	\\
	&	 HR@5 	& NDCG@5 &	 HR@10 	& NDCG@10 	\\
	%\midrule
\cmidrule(lr){2-5}

%    Random	&0.0084	&0.0084	&0.0510	&0.0290	&0.0987	&0.0442	\\
    TopPopular	&0.3043	&0.2062	&0.4531	&0.2542	\\
    UserKNN	&0.4916	&0.3328	&0.6705	&0.3908	\\
    ItemKNN	&0.4829	&0.3328	&0.6596	&0.3900	\\
    \smash{\palpha}	&0.4811	&0.3331	&0.6464	&0.3867	\\
    \smash{\pbeta}		&0.4922	&0.3409	&0.6715	&0.3991	\\
    \midrule
    NeuMF		&0.5486	&0.3840	&0.7120	&0.4369	\\ 
    \midrule
    SLIM	&\textbf{0.5589}	&\textbf{0.3961}	&\textbf{0.7161}	&\textbf{0.4470}	\\
 	\bottomrule
  	\end{tabular}
%  	}
\end{table}

Given the publicly shared information, we could reproduce the results from \cite{he2017neural}. The outcomes of the experiment are shown in Table \ref{tab:NCF_results}. 
On the Pinterest dataset, two of the personalized baselines were better than NeuMF on all metrics. 
For the MovieLens dataset, NeuMF outperformed our simple baselines quite clearly.

Since the MovieLens dataset has been extensively used over the last decades for evaluating new models, we made additional experiments with \emph{SLIM}, a simple linear method described in \cite{ning2011SLIM}. 
To implement \emph{SLIM}, we took the standard Elastic Net implementation provided in the \small{\texttt{scikit-learn}} \normalsize package for Python (\small{\texttt{ElasticNet}}\normalsize).
To tune the hyper-parameters on the validation set, we considered neighborhood sizes as in the other baselines; the ratio of l1 and l2 regularization between $10^{-5}$ and $1.0$; and the regularization magnitude coefficient between $10^{-3}$ and $1.0$. 
Table \ref{tab:NCF_results} shows that \emph{SLIM} is indeed better than our baselines, as expected, but also outperforms NeuMF on this dataset. 

\subsection{Spectral Collaborative Filtering (SpectralCF)}
\label{subsec:scf}
SpectralCF \cite{Zheng:2018:SCF:3240323.3240343}, presented at RecSys '18, was designed to specifically address the cold-start problem and is based on concepts of Spectral Graph Theory. Its recommendations are based on the bipartite user-item relationship graph and a novel convolution operation, which is used to make collaborative recommendations directly in the \emph{spectral domain}.
The method was evaluated on three public datasets (MovieLens1M, HetRec, and Amazon Instant Video) and benchmarked against a variety of methods, including recent neural approaches and established factorization and ranking techniques. The evaluation was based on randomly created 80/20 training-test splits and using Recall and the Mean Average Precision (MAP) at different cutoffs.\footnote{To assess the cold-start behavior, additional experiments are performed with fewer data points per user in the training set.}

For the MovieLens dataset, the training and test datasets used by the authors were shared along with the code. For the other datasets, the data splits were not published therefore we created the splits by ourself following the descriptions in the paper.

Somehow surprisingly, the authors report only one set of hyper-parameter values in the paper, which they apparently used for all datasets. We therefore ran the code both with the provided hyper-parameters and with hyper-parameter settings that we determined by our own on all datasets. For the HetRec and Amazon Instant Video datasets, all our baselines, to our surprise also including the TopPoular method, outperformed SpectralCF on all measures. However, when running the code on the provided MovieLens data splits, we found that SpectralCF was better than all our baselines by a huge margin. Recall@20 was, for example, 50\% higher than our best baseline.

We therefore analyzed the published train-test split for the MovieLens dataset and observed that the popularity distribution of the items in the test set is very different from a distribution that would likely result from a random sampling procedure.\footnote{We contacted the authors on this issue, but did not receive an explanation for this phenomenon.} We then ran experiments with our own train-test splits also for the MovieLens dataset, using the splitting procedure described in the paper. We optimized the parameters for our data split to ensure a fair comparison.
The results of the experiment are shown in Table \ref{tab:SpectralCF_results}. When using data splits that were created as described in the original paper, the results for the MovieLens dataset are in line with our own experiments for the other two datasets, i.e., SpectralCF in all configurations performed worse than our baseline methods and was outperformed even by the TopPopular method.

\begin{table}[h!t]
    \caption{Experimental results for SpectralCF (MovieLens1M, using own random splits and five repeated measurements).}
    \label{tab:SpectralCF_results}
\resizebox{\linewidth}{!}{%
    \begin{tabular}{@{}lcccccc@{}}
    \toprule
	  & \multicolumn{2}{c}{Cutoff 20} 	& \multicolumn{2}{c}{Cutoff 60} 	& \multicolumn{2}{c}{Cutoff 100} 	\\
	  & REC 	& MAP 	& REC 	& MAP 	& REC 	& MAP 	\\
		\midrule
%    Random	&0.0052	&0.0010	&0.0162	&0.0013		&0.0255	&0.0015	\\
    TopPopular	&\textbf{0.1853}	&\textbf{0.0576}	&\textbf{0.3335}	&\textbf{0.0659}		&\textbf{0.4244}	&\textbf{0.0696}	\\
    UserKNN CF	&\textbf{0.2881}	&\textbf{0.1106}		&\textbf{0.4780}	&\textbf{0.1238}		&\textbf{0.5790}	&\textbf{0.1290}	\\
    ItemKNN CF	&\textbf{0.2819}	&\textbf{0.1059}		&\textbf{0.4712}	&\textbf{0.1190}		&\textbf{0.5737}	&\textbf{0.1243}	\\
    \smash{\palpha}	&\textbf{0.2853}	&\textbf{0.1051}		&\textbf{0.4808}	&\textbf{0.1195}		&\textbf{0.5760}	&\textbf{0.1248}	\\
    \smash{\pbeta}	&\textbf{0.2910}	&\textbf{0.1088}		&\textbf{0.4882}	&\textbf{0.1233}		&\textbf{0.5884}	&\textbf{0.1288}	\\ \midrule
%    PureSVD	&\textbf{0.2522}	&\textbf{0.0918}		&\textbf{0.4318}	&\textbf{0.1039}		&\textbf{0.5288}	&\textbf{0.1089}	\\ \midrule
    %SpectralCF	&0.1808	&0.0559	&0.3246	&0.0636	&0.4196	&0.0672	\\
    SpectralCF	&0.1843	&0.0539	&0.3274	&0.0618	&0.4254	&0.0656	\\
	\bottomrule
  	\end{tabular}
 	}
\end{table}

\begin{figure}[h!t]
\centering
\resizebox{\linewidth}{!}{%
\input{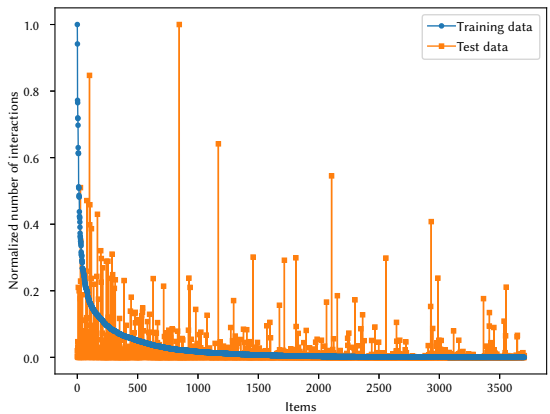}
}
 \caption{Popularity distributions of the provided training and test splits. In case of a random split, the normalized values should, on average, be close for both splits.}
 \label{fig:SpectralCF_dataset_popularity_distribution_ours}
\end{figure}

Figure \ref{fig:SpectralCF_dataset_popularity_distribution_ours} visualizes the data splitting problem. The blue data points show the normalized popularity values for each item in the training set, with the most popular item in the corresponding split having the value 1, ordered by decreasing popularity values. In case of random sampling of ratings, the orange points from the test set would mostly be very close to the corresponding blue ones. Here, however, we see that the popularity values of many items in the test set differ largely. An analysis of the distributions with measures like the Gini index or Shannon entropy confirms that the dataset characteristics of the shared test set diverge largely from a random split. The Gini index of a true random split lies at around 0.79 for both the training and test split. While the Gini index for the provided training split is similar to ours, the Gini index of the provided test split is much higher (0.92), which means that the distribution has a much higher popularity bias than a random split.

\subsection{Variational Autoencoders for Collaborative Filtering (Mult-VAE)} % 2018
\label{subsec:vacf}
Mult-VAE \cite{liang2018variationalautoencodersforCF} is a collaborative filtering method for implicit feedback based on variational autoencoders. The work was presented at WWW '18. With Mult-VAE, the authors introduce a generative model with multinomial likelihood, propose a different regularization parameter for the learning objective, and use Bayesian inference for parameter estimation. They evaluate their method on three binarized datasets that originally contain movie ratings or song play counts. The baselines in the experiments include both a matrix factorization method from 2008 \cite{HuKorenVolinsky2008}, a linear model from 2011 \cite{ning2011SLIM}, and a more recent neural method \cite{wu2016collaborative}. Accoring to the reported experiments, the proposed method leads to accuracy results that are typically around 3\% better than the best baseline in terms of Recall and the NDCG.

Using their code and datasets, we found that the proposed method indeed consistently outperforms our quite simple baseline techniques. The obtained accuracy results were between 10\% and 20\% better than our best baseline.
Thus, with Mult-VAE, we found one example in the examined literature where a more complex method was better, by a large margin, than any of our baseline techniques in all configurations.

To validate that Mult-VAE is advantageous over the complex non-neural models, as reported in \cite{liang2018variationalautoencodersforCF}, we optimized the parameters for the weighted matrix factorization technique \cite{HuKorenVolinsky2008} and the linear model \cite{ning2011SLIM} (SLIM using Elastic Net) for the MovieLens and Netflix datasets by ourselves. We made the following observations. For both datasets, we could reproduce the results and observe improvements over SLIM of up to 5\% on the different measures reported in the original papers. Table \ref{tab:MultVAE-results-netflix-original} shows the outcomes for the Netflix datasets using the measurements and cutoffs from the original experiments after optimizing for NDCG@100 as in \cite{liang2018variationalautoencodersforCF}.

\begin{table}[h!t]
    \caption{Experimental results for Mult-VAE (Netflix data), using metrics and cutoffs reported in the original paper.}
    \label{tab:MultVAE-results-netflix-original}
    \begin{tabular}{lccc}
    \toprule
		& REC@20 	& REC@50 	& NDCG@100 	\\
    \midrule
    TopPop	&0.0782		&0.1643	&0.1570	\\
    ItemKNN CF	&0.2088		&0.3386		&0.3086	\\
    \smash{\palpha}     	&0.1977		&0.3346		&0.2967	\\
    \smash{\pbeta}	&0.2196		&0.3560		&0.3246	\\
    \midrule
    SLIM	&0.2551		&0.3995		&0.3745	\\
    \midrule
    Mult-VAE &\textbf{0.2626}		&\textbf{0.4138}	&\textbf{0.3756}	\\
	\bottomrule
  	\end{tabular}
\end{table}

The differences between Mult-VAE and SLIM in terms of the NDCG, the optimization goal, are quite small. In terms of the Recall, however, Mult-VAE improvements over SLIM seem solid. Since the choice of the used cutoffs (20 and 50 for Recall, and 100 for NDCG) is not very consistent in \cite{liang2018variationalautoencodersforCF}, we made additional measurements at different cutoff lengths. The results are provided in Table \ref{tab:MultVAE-results-netflix-additional}. They show that when using the NDCG as an optimization goal \emph{and} as a performance measure, the differences between SLIM and Mult-VAE disappear on this dataset, and SLIM is actually sometimes slightly better. A similar phenomenon can be observed for the MovieLens dataset. In this particular case, therefore, the progress that is achieved through the neural approach is only partial and depends on the chosen evaluation measure.

\begin{table}[h!t]
    \caption{Experimental results for Mult-VAE using additional cutoff lengths for the Netflix dataset.}
    \label{tab:MultVAE-results-netflix-additional}
\resizebox{\linewidth}{!}{%
    \begin{tabular}{lcccc}
    \toprule
		&  NDCG@20 	& NDCG@50 	& REC@100 	& NDCG@100 	\\ \midrule
%    TopPop		&0.0761		&0.1159	&0.2718	&0.1570	\\
%    ItemKNN CF	&0.1966		&0.2588	&0.4590	&0.3086	\\
%    P3alpha		&0.1762		&0.2414	&0.4675	&0.2967	\\
%    RP3beta		&0.2044		&0.2699	&0.4886	&0.3246	\\
%    \midrule
    SLIM		&\textbf{0.2473}	&\textbf{0.3196}	&0.5289	&0.3745	\\
    \midrule
    Mult-VAE 	&0.2448	& 0.3192	&\textbf{0.5476}	&\textbf{0.3756}	\\
	\bottomrule
  	\end{tabular}
  	}
\end{table}

\section{Discussion}
\label{sec:discussion}
\subsection{Reproducibility and Scalability}
In some ways, establishing reproducibility in applied machine learning should be much easier than in other scientific disciplines and also other subfields of computer science. While many recommendation algorithms are not fully deterministic, e.g., because they use some form of random initialization of parameters, the variability of the obtained results when repeating the exact same experiment configuration several times is probably very low in most cases. Therefore, when researchers provide their code and the used data, everyone should be able to reproduce more or less the exact same results. Given that researchers today often rely on software that is publicly available or provided by academic institutions, the barriers regarding technological requirements are mostly low as well.
In particular, virtualization technology should make it easier for other researchers to repeat an experiment under very similar conditions.

Nonetheless, our work shows that the level of reproducibility is actually not high. The code of the core algorithms seems to be more often shared by researchers than in the past, probably also due to the fact that reproducibility has become an evaluation criterion for conferences. However, in many cases, the code that is used for hyper-parameter optimization, evaluation, data pre-processing, and for the baselines is not shared.
This makes it difficult for others to validate the reported findings.

One orthogonal factor that can make reproducibility challenging is the computational complexity of many of the proposed methods. Ten years after the Netflix Prize and its 100 million rating dataset, researchers, in the year 2019, commonly use datasets containing only a few hundred thousand ratings. Even for such tiny datasets, which were considered unacceptably small a few years ago, hyper-parameter optimization can take days or weeks, even when researchers have access to GPU computing. Clearly, nearest-neighbor methods, as discussed in our paper, can also lead to scalability issues. However, with appropriate data pre-processing and data sampling mechanisms, scalability can also be ensured for such methods, both in academic and industrial environments \cite{Linden2003,JannachLudewig2017RecSys}.

\subsection{Progress Assessment}
Despite their computational complexity, our analysis showed that several recently proposed neural methods do not even outperform conceptually or computationally simpler, sometimes long-known, algorithms. The level of progress that is achieved in the field of neural methods is, therefore, unclear, at least when considering the approaches discussed in our paper.

One main reason for this \emph{phantom progress}, as our work shows, lies in the choice of the baselines and the lack of a proper optimization of the baselines. In the majority of the investigated cases, not enough information is given about the optimization of the considered baselines. Sometimes, we also found that mistakes were made with respect to data splitting and the implementation of certain evaluation measures and protocols.

Another interesting observation is that a number of recent papers use the neural collaborative filtering method (NCF) \cite{he2017neural} as one of their state-of-the-art baselines. According to our analysis, this method is however outperformed by simple baselines on one dataset and does not lead to much better results on another, where it is also outperformed by a standard implementation of a linear regression method. Therefore, progress is often claimed by comparing a complex neural model against another neural model, which is, however, not necessarily a strong baseline. Similar observations can be made for the area of session-based recommendation, where a recent method based on recurrent neural networks \cite{DBLP:journals/corr/HidasiKBT15} is considered a competitive baseline, even though almost trivial methods are in most cases better \cite{Ludewig2018,LudewigMauro2019}.

Another aspect that makes it difficult to assess progress in the field lies in the variety of datasets, evaluation protocols, metrics, and baselines that are used by researchers. Regarding datasets, for example, we found over 20 public datasets that were used, plus several variants of the MovieLens and Yelp datasets. As a result, most datasets are only used in one or two papers. All sorts of metrics are used (e.g., Precision, Recall, Mean Average Precision, NDCG, MRR etc.) as well as various evaluation procedures (e.g., random holdout 80/20, leave-last-out, leave-one-out, 100 negative items or 50 negative items for each positive). In most cases, however, these choices are not well justified beyond the fact that others used them before. In reality, the choice of the metric should depend on the application context. In some applications, for example, it might be important to have at least one relevant item at the top of the recommendations, which suggests the use of rank-based metrics like MRR. In other domains, high Recall might be more important when the goal is to show as many relevant items as possible to the user. Besides the unclear choice of the measure, often also the cutoff sizes for the measurement are not explained and range from top-3 or top-5 lists to several hundred elements.

These phenomena are, however, not tied to neural recommendation approaches, but can be found in algorithmic research in recommender systems also in pre-neural times. Considering the arguments from \cite{troubling-trends-1807.03341,DBLP:journals/corr/abs-1206-4656}, such developments are fueled by the strong focus of machine learning researchers on accuracy measures and the hunt for the ``best'' model. In our current research practice, it is often considered sufficient to show that a new method can outperform a set of existing algorithms on at least one or two public datasets on one or two established accuracy measures.\footnote{From the 18 papers considered relevant for our study, there were at least two papers which proposed new DL architectures which were evaluated on a single private dataset and for which no source code was provided.}
The choice of the evaluation measure and dataset however often seems arbitrary.

An example of such unclear research practice is the use of MovieLens rating datasets for the evaluation of algorithms for implicit feedback datasets. Such practices point to the underlying fundamental problem that research is not guided by any hypothesis or aim at the solution of a given problem. The hunt for better accuracy values dominates research activities in this area, even though it is not even clear if slightly higher accuracy values are relevant in terms of adding value for recommendation consumers or providers \cite{Konstan2012,Xiao:2007:EPR:2017327.2017335,JannachResnickEtAl2016}. In fact, a number of research works exist that indicate that higher accuracy does not necessarily translate into better-received recommendations \cite{Rossetti:2016:COO:2959100.2959176,Cremonesi:2012:IPP:2209310.2209314,Garcin:2014:OOE:2645710.2645745,Maksai:2015:POP:2792838.2800184,DBLP:conf/ercimdl/BeelL15}.

\section{Summary}
\label{sec:summary}
In this work, we have analyzed a number of recent neural algorithms for top-n recommendation.
Our analysis indicates that reproducing published research is still challenging. Furthermore, it turned out that most of the reviewed works can be outperformed at least on some datasets by conceptually and computationally simpler algorithms. Our work therefore calls for more rigor and better research practices with respect to the evaluation of algorithmic contributions in this area.

Our analyses so far are limited to papers published in certain conference series. In our ongoing and future work, we plan to extend our analysis to other publication outlets and other types of recommendation problems. Furthermore, we plan to consider more traditional algorithms as baselines, e.g., based on matrix factorization.

\bibliographystyle{ACM-Reference-Format}
\bibliography{main-references}


\begin{thebibliography}{53}

%%% the Web address.

\ifx \showCODEN    \undefined \def \showCODEN     #1{\unskip}     \fi
\ifx \showDOI      \undefined \def \showDOI       #1{#1}\fi
\ifx \showISBNx    \undefined \def \showISBNx     #1{\unskip}     \fi
\ifx \showISBNxiii \undefined \def \showISBNxiii  #1{\unskip}     \fi
\ifx \showISSN     \undefined \def \showISSN      #1{\unskip}     \fi
\ifx \showLCCN     \undefined \def \showLCCN      #1{\unskip}     \fi
\ifx \shownote     \undefined \def \shownote      #1{#1}          \fi
\ifx \showarticletitle \undefined \def \showarticletitle #1{#1}   \fi
\ifx \showURL      \undefined \def \showURL       {\relax}        \fi
% invisible to TeX
\providecommand\bibfield[2]{#2}
\providecommand\bibinfo[2]{#2}
\providecommand\natexlab[1]{#1}
\providecommand\showeprint[2][]{arXiv:#2}

\bibitem[\protect\citeauthoryear{Antenucci, Boglio, Chioso, Dervishaj, Shuwen,
  Scarlatti, and Ferrari~Dacrema}{Antenucci et~al\mbox{.}}{2018}]%
        {antenucci2018artist}
\bibfield{author}{\bibinfo{person}{S. Antenucci}, \bibinfo{person}{S. Boglio},
  \bibinfo{person}{E. Chioso}, \bibinfo{person}{E. Dervishaj},
  \bibinfo{person}{K. Shuwen}, \bibinfo{person}{T. Scarlatti}, {and}
  \bibinfo{person}{M. Ferrari~Dacrema}.} \bibinfo{year}{2018}\natexlab{}.
\newblock \showarticletitle{Artist-driven layering and user's behaviour impact
  on recommendations in a playlist continuation scenario}. In
  \bibinfo{booktitle}{\emph{Proceedings of the ACM Recommender Systems
  Challenge 2018 (RecSys 2018)}}.
\newblock
\urldef\tempurl%
\url{https://doi.org/10.1145/3267471.3267475}
\showDOI{\tempurl}
\newblock
\shownote{Source:
  \url{https://github.com/MaurizioFD/spotify-recsys-challenge}.}


\bibitem[\protect\citeauthoryear{Armstrong, Moffat, Webber, and
  Zobel}{Armstrong et~al\mbox{.}}{2009}]%
        {Armstrong:2009:IDA:1645953.1646031}
\bibfield{author}{\bibinfo{person}{Timothy~G. Armstrong},
  \bibinfo{person}{Alistair Moffat}, \bibinfo{person}{William Webber}, {and}
  \bibinfo{person}{Justin Zobel}.} \bibinfo{year}{2009}\natexlab{}.
\newblock \showarticletitle{Improvements That Don't Add Up: Ad-hoc Retrieval
  Results Since 1998}. In \bibinfo{booktitle}{\emph{Proceedings CIKM '09}}.
  \bibinfo{pages}{601--610}.
\newblock


\bibitem[\protect\citeauthoryear{Beel, Breitinger, Langer, Lommatzsch, and
  Gipp}{Beel et~al\mbox{.}}{2016}]%
        {Beel2016}
\bibfield{author}{\bibinfo{person}{Joeran Beel}, \bibinfo{person}{Corinna
  Breitinger}, \bibinfo{person}{Stefan Langer}, \bibinfo{person}{Andreas
  Lommatzsch}, {and} \bibinfo{person}{Bela Gipp}.}
  \bibinfo{year}{2016}\natexlab{}.
\newblock \showarticletitle{Towards reproducibility in recommender-systems
  research}.
\newblock \bibinfo{journal}{\emph{User Modeling and User-Adapted Interaction}}
  \bibinfo{volume}{26}, \bibinfo{number}{1} (\bibinfo{year}{2016}),
  \bibinfo{pages}{69--101}.
\newblock


\bibitem[\protect\citeauthoryear{Beel and Langer}{Beel and Langer}{2015}]%
        {DBLP:conf/ercimdl/BeelL15}
\bibfield{author}{\bibinfo{person}{J{\"{o}}ran Beel} {and}
  \bibinfo{person}{Stefan Langer}.} \bibinfo{year}{2015}\natexlab{}.
\newblock \showarticletitle{A Comparison of Offline Evaluations, Online
  Evaluations, and User Studies in the Context of Research-Paper Recommender
  Systems}. In \bibinfo{booktitle}{\emph{Proceedings TPDL '15}}.
  \bibinfo{pages}{153--168}.
\newblock


\bibitem[\protect\citeauthoryear{Bell and Koren}{Bell and Koren}{2007}]%
        {bell2007improved}
\bibfield{author}{\bibinfo{person}{Robert~M Bell} {and} \bibinfo{person}{Yehuda
  Koren}.} \bibinfo{year}{2007}\natexlab{}.
\newblock \showarticletitle{Improved neighborhood-based collaborative
  filtering}. In \bibinfo{booktitle}{\emph{KDD cup and workshop at the KDD
  '07}}. Citeseer, \bibinfo{pages}{7--14}.
\newblock


\bibitem[\protect\citeauthoryear{Bharadhwaj, Park, and Lim}{Bharadhwaj
  et~al\mbox{.}}{2018}]%
        {Bharadhwaj:2018:RRG:3240323.3240383}
\bibfield{author}{\bibinfo{person}{Homanga Bharadhwaj}, \bibinfo{person}{Homin
  Park}, {and} \bibinfo{person}{Brian~Y. Lim}.}
  \bibinfo{year}{2018}\natexlab{}.
\newblock \showarticletitle{RecGAN: Recurrent Generative Adversarial Networks
  for Recommendation Systems}. In \bibinfo{booktitle}{\emph{Proceedings RecSys
  '18}}. \bibinfo{pages}{372--376}.
\newblock


\bibitem[\protect\citeauthoryear{Chen, Zhang, He, Nie, Liu, and Chua}{Chen
  et~al\mbox{.}}{2017}]%
        {chen2017attentivecomponentlevelattention}
\bibfield{author}{\bibinfo{person}{Jingyuan Chen}, \bibinfo{person}{Hanwang
  Zhang}, \bibinfo{person}{Xiangnan He}, \bibinfo{person}{Liqiang Nie},
  \bibinfo{person}{Wei Liu}, {and} \bibinfo{person}{Tat-Seng Chua}.}
  \bibinfo{year}{2017}\natexlab{}.
\newblock \showarticletitle{Attentive collaborative filtering: Multimedia
  recommendation with item-and component-level attention}. In
  \bibinfo{booktitle}{\emph{Proceedings SIGIR '17}}. \bibinfo{pages}{335--344}.
\newblock


\bibitem[\protect\citeauthoryear{Cooper, Lee, Radzik, and Siantos}{Cooper
  et~al\mbox{.}}{2014}]%
        {cooper2014P3alpha}
\bibfield{author}{\bibinfo{person}{Colin Cooper}, \bibinfo{person}{Sang~Hyuk
  Lee}, \bibinfo{person}{Tomasz Radzik}, {and} \bibinfo{person}{Yiannis
  Siantos}.} \bibinfo{year}{2014}\natexlab{}.
\newblock \showarticletitle{Random walks in recommender systems: exact
  computation and simulations}. In \bibinfo{booktitle}{\emph{Proceedings WWW
  '14}}. \bibinfo{pages}{811--816}.
\newblock


\bibitem[\protect\citeauthoryear{Cremonesi, Garzotto, and Turrin}{Cremonesi
  et~al\mbox{.}}{2012}]%
        {Cremonesi:2012:IPP:2209310.2209314}
\bibfield{author}{\bibinfo{person}{Paolo Cremonesi}, \bibinfo{person}{Franca
  Garzotto}, {and} \bibinfo{person}{Roberto Turrin}.}
  \bibinfo{year}{2012}\natexlab{}.
\newblock \showarticletitle{Investigating the Persuasion Potential of
  Recommender Systems from a Quality Perspective: An Empirical Study}.
\newblock \bibinfo{journal}{\emph{Transactions on Interactive Intelligent
  Systems}} \bibinfo{volume}{2}, \bibinfo{number}{2} (\bibinfo{year}{2012}),
  \bibinfo{pages}{1--41}.
\newblock


\bibitem[\protect\citeauthoryear{Ebesu, Shen, and Fang}{Ebesu
  et~al\mbox{.}}{2018}]%
        {ebesu2018collaborative}
\bibfield{author}{\bibinfo{person}{Travis Ebesu}, \bibinfo{person}{Bin Shen},
  {and} \bibinfo{person}{Yi Fang}.} \bibinfo{year}{2018}\natexlab{}.
\newblock \showarticletitle{Collaborative Memory Network for Recommendation
  Systems}. In \bibinfo{booktitle}{\emph{Proceedings SIGIR '18}}.
  \bibinfo{pages}{515--524}.
\newblock


\bibitem[\protect\citeauthoryear{Elkahky, Song, and He}{Elkahky
  et~al\mbox{.}}{2015}]%
        {elkahky2015multiviewdeeplearningcrossdomain}
\bibfield{author}{\bibinfo{person}{Ali~Mamdouh Elkahky}, \bibinfo{person}{Yang
  Song}, {and} \bibinfo{person}{Xiaodong He}.} \bibinfo{year}{2015}\natexlab{}.
\newblock \showarticletitle{A multi-view deep learning approach for cross
  domain user modeling in recommendation systems}. In
  \bibinfo{booktitle}{\emph{Proceedings WWW '15}}. \bibinfo{pages}{278--288}.
\newblock


\bibitem[\protect\citeauthoryear{for Computing~Machinery}{for
  Computing~Machinery}{2016}]%
        {ACMreproducibilty2016}
\bibfield{author}{\bibinfo{person}{Association for Computing~Machinery}.}
  \bibinfo{year}{2016}\natexlab{}.
\newblock \bibinfo{title}{Artifact Review and Badging}.
\newblock \bibinfo{howpublished}{Available online at:
  https://www.acm.org/publications/policies/artifact-review-badging (Accessed
  March, 2018)}.
\newblock


\bibitem[\protect\citeauthoryear{Garcin, Faltings, Donatsch, Alazzawi, Bruttin,
  and Huber}{Garcin et~al\mbox{.}}{2014}]%
        {Garcin:2014:OOE:2645710.2645745}
\bibfield{author}{\bibinfo{person}{Florent Garcin}, \bibinfo{person}{Boi
  Faltings}, \bibinfo{person}{Olivier Donatsch}, \bibinfo{person}{Ayar
  Alazzawi}, \bibinfo{person}{Christophe Bruttin}, {and} \bibinfo{person}{Amr
  Huber}.} \bibinfo{year}{2014}\natexlab{}.
\newblock \showarticletitle{Offline and Online Evaluation of News Recommender
  Systems at {Swissinfo.Ch}}. In \bibinfo{booktitle}{\emph{Proceedings RecSys
  '14}}. \bibinfo{pages}{169--176}.
\newblock


\bibitem[\protect\citeauthoryear{He, Liao, Zhang, Nie, Hu, and Chua}{He
  et~al\mbox{.}}{2017}]%
        {he2017neural}
\bibfield{author}{\bibinfo{person}{Xiangnan He}, \bibinfo{person}{Lizi Liao},
  \bibinfo{person}{Hanwang Zhang}, \bibinfo{person}{Liqiang Nie},
  \bibinfo{person}{Xia Hu}, {and} \bibinfo{person}{Tat-Seng Chua}.}
  \bibinfo{year}{2017}\natexlab{}.
\newblock \showarticletitle{Neural collaborative filtering}. In
  \bibinfo{booktitle}{\emph{Proceedings WWW '17}}. \bibinfo{pages}{173--182}.
\newblock


\bibitem[\protect\citeauthoryear{Henderson, Islam, Bachman, Pineau, Precup, and
  Meger}{Henderson et~al\mbox{.}}{2018}]%
        {DBLP:conf/aaai/0002IBPPM18}
\bibfield{author}{\bibinfo{person}{Peter Henderson}, \bibinfo{person}{Riashat
  Islam}, \bibinfo{person}{Philip Bachman}, \bibinfo{person}{Joelle Pineau},
  \bibinfo{person}{Doina Precup}, {and} \bibinfo{person}{David Meger}.}
  \bibinfo{year}{2018}\natexlab{}.
\newblock \showarticletitle{Deep Reinforcement Learning That Matters}. In
  \bibinfo{booktitle}{\emph{Proceedings AAAI '18}}.
  \bibinfo{pages}{3207--3214}.
\newblock


\bibitem[\protect\citeauthoryear{Hidasi, Karatzoglou, Baltrunas, and
  Tikk}{Hidasi et~al\mbox{.}}{2016}]%
        {DBLP:journals/corr/HidasiKBT15}
\bibfield{author}{\bibinfo{person}{Bal{\'{a}}zs Hidasi},
  \bibinfo{person}{Alexandros Karatzoglou}, \bibinfo{person}{Linas Baltrunas},
  {and} \bibinfo{person}{Domonkos Tikk}.} \bibinfo{year}{2016}\natexlab{}.
\newblock \showarticletitle{Session-based Recommendations with Recurrent Neural
  Networks}. In \bibinfo{booktitle}{\emph{Proceedings ICLR '16}}.
\newblock


\bibitem[\protect\citeauthoryear{Hu, Shi, Zhao, and Yu}{Hu
  et~al\mbox{.}}{2018}]%
        {hu2018leveragingmetapathcontext}
\bibfield{author}{\bibinfo{person}{Binbin Hu}, \bibinfo{person}{Chuan Shi},
  \bibinfo{person}{Wayne~Xin Zhao}, {and} \bibinfo{person}{Philip~S Yu}.}
  \bibinfo{year}{2018}\natexlab{}.
\newblock \showarticletitle{Leveraging meta-path based context for top-n
  recommendation with a neural co-attention model}. In
  \bibinfo{booktitle}{\emph{Proceedings KDD '18}}. \bibinfo{pages}{1531--1540}.
\newblock


\bibitem[\protect\citeauthoryear{{Hu}, {Koren}, and {Volinsky}}{{Hu}
  et~al\mbox{.}}{2008}]%
        {HuKorenVolinsky2008}
\bibfield{author}{\bibinfo{person}{Yifan {Hu}}, \bibinfo{person}{Yehuda
  {Koren}}, {and} \bibinfo{person}{Chris {Volinsky}}.}
  \bibinfo{year}{2008}\natexlab{}.
\newblock \showarticletitle{Collaborative Filtering for Implicit Feedback
  Datasets}. In \bibinfo{booktitle}{\emph{Proceedings ICDM '08}}.
  \bibinfo{pages}{263--272}.
\newblock


\bibitem[\protect\citeauthoryear{Jannach and Ludewig}{Jannach and
  Ludewig}{2017}]%
        {JannachLudewig2017RecSys}
\bibfield{author}{\bibinfo{person}{Dietmar Jannach} {and}
  \bibinfo{person}{Malte Ludewig}.} \bibinfo{year}{2017}\natexlab{}.
\newblock \showarticletitle{When Recurrent Neural Networks Meet the
  Neighborhood for Session-Based Recommendation}. In
  \bibinfo{booktitle}{\emph{Proceedings RecSys '17}}.
  \bibinfo{pages}{306--310}.
\newblock


\bibitem[\protect\citeauthoryear{Jannach, Resnick, Tuzhilin, and
  Zanker}{Jannach et~al\mbox{.}}{2016}]%
        {JannachResnickEtAl2016}
\bibfield{author}{\bibinfo{person}{Dietmar Jannach}, \bibinfo{person}{Paul
  Resnick}, \bibinfo{person}{Alexander Tuzhilin}, {and} \bibinfo{person}{Markus
  Zanker}.} \bibinfo{year}{2016}\natexlab{}.
\newblock \showarticletitle{Recommender Systems - Beyond Matrix Completion}.
\newblock \bibinfo{journal}{\emph{Commun. ACM}} \bibinfo{volume}{59},
  \bibinfo{number}{11} (\bibinfo{year}{2016}), \bibinfo{pages}{94--102}.
\newblock


\bibitem[\protect\citeauthoryear{Kim, Park, Oh, Lee, and Yu}{Kim
  et~al\mbox{.}}{2016}]%
        {Kim:2016:CMF:2959100.2959165}
\bibfield{author}{\bibinfo{person}{Donghyun Kim}, \bibinfo{person}{Chanyoung
  Park}, \bibinfo{person}{Jinoh Oh}, \bibinfo{person}{Sungyoung Lee}, {and}
  \bibinfo{person}{Hwanjo Yu}.} \bibinfo{year}{2016}\natexlab{}.
\newblock \showarticletitle{Convolutional Matrix Factorization for Document
  Context-Aware Recommendation}. In \bibinfo{booktitle}{\emph{Proceedings
  RecSys '16}}. \bibinfo{pages}{233--240}.
\newblock


\bibitem[\protect\citeauthoryear{Konstan and Riedl}{Konstan and Riedl}{2012}]%
        {Konstan2012}
\bibfield{author}{\bibinfo{person}{Joseph~A. Konstan} {and}
  \bibinfo{person}{John Riedl}.} \bibinfo{year}{2012}\natexlab{}.
\newblock \showarticletitle{Recommender systems: from algorithms to user
  experience}.
\newblock \bibinfo{journal}{\emph{User Modeling and User-Adapted Interaction}}
  \bibinfo{volume}{22}, \bibinfo{number}{1} (\bibinfo{year}{2012}),
  \bibinfo{pages}{101--123}.
\newblock
\showISSN{1573-1391}


\bibitem[\protect\citeauthoryear{Li and She}{Li and She}{2017}]%
        {li2017collaborativevariationalautoencoder}
\bibfield{author}{\bibinfo{person}{Xiaopeng Li} {and} \bibinfo{person}{James
  She}.} \bibinfo{year}{2017}\natexlab{}.
\newblock \showarticletitle{Collaborative variational autoencoder for
  recommender systems}. In \bibinfo{booktitle}{\emph{Proceedings KDD '17}}.
  \bibinfo{pages}{305--314}.
\newblock


\bibitem[\protect\citeauthoryear{Liang, Krishnan, Hoffman, and Jebara}{Liang
  et~al\mbox{.}}{2018}]%
        {liang2018variationalautoencodersforCF}
\bibfield{author}{\bibinfo{person}{Dawen Liang}, \bibinfo{person}{Rahul~G
  Krishnan}, \bibinfo{person}{Matthew~D Hoffman}, {and} \bibinfo{person}{Tony
  Jebara}.} \bibinfo{year}{2018}\natexlab{}.
\newblock \showarticletitle{Variational Autoencoders for Collaborative
  Filtering}. In \bibinfo{booktitle}{\emph{Proceedings WWW '18}}.
  \bibinfo{pages}{689--698}.
\newblock


\bibitem[\protect\citeauthoryear{Lin}{Lin}{2019}]%
        {Lin:2019:NHC:3308774.3308781}
\bibfield{author}{\bibinfo{person}{Jimmy Lin}.}
  \bibinfo{year}{2019}\natexlab{}.
\newblock \showarticletitle{The Neural Hype and Comparisons Against Weak
  Baselines}.
\newblock \bibinfo{journal}{\emph{SIGIR Forum}} \bibinfo{volume}{52},
  \bibinfo{number}{2} (\bibinfo{date}{Jan.} \bibinfo{year}{2019}),
  \bibinfo{pages}{40--51}.
\newblock


\bibitem[\protect\citeauthoryear{{Linden}, {Smith}, and {York}}{{Linden}
  et~al\mbox{.}}{2003}]%
        {Linden2003}
\bibfield{author}{\bibinfo{person}{G. {Linden}}, \bibinfo{person}{B. {Smith}},
  {and} \bibinfo{person}{J. {York}}.} \bibinfo{year}{2003}\natexlab{}.
\newblock \showarticletitle{Amazon.com recommendations: item-to-item
  collaborative filtering}.
\newblock \bibinfo{journal}{\emph{IEEE Internet Computing}}
  \bibinfo{volume}{7}, \bibinfo{number}{1} (\bibinfo{year}{2003}),
  \bibinfo{pages}{76--80}.
\newblock


\bibitem[\protect\citeauthoryear{Lipton and Steinhardt}{Lipton and
  Steinhardt}{2018}]%
        {troubling-trends-1807.03341}
\bibfield{author}{\bibinfo{person}{Zachary~C. Lipton} {and}
  \bibinfo{person}{Jacob Steinhardt}.} \bibinfo{year}{2018}\natexlab{}.
\newblock \bibinfo{title}{Troubling Trends in Machine Learning Scholarship}.
\newblock
\newblock
\showeprint{arXiv:1807.03341}


\bibitem[\protect\citeauthoryear{Lops, De~Gemmis, and Semeraro}{Lops
  et~al\mbox{.}}{2011}]%
        {lops2011content}
\bibfield{author}{\bibinfo{person}{Pasquale Lops}, \bibinfo{person}{Marco
  De~Gemmis}, {and} \bibinfo{person}{Giovanni Semeraro}.}
  \bibinfo{year}{2011}\natexlab{}.
\newblock \showarticletitle{Content-based recommender systems: State of the art
  and trends}.
\newblock In \bibinfo{booktitle}{\emph{Recommender Systems Handbook}}.
  \bibinfo{publisher}{Springer}, \bibinfo{pages}{73--105}.
\newblock


\bibitem[\protect\citeauthoryear{Ludewig and Jannach}{Ludewig and
  Jannach}{2018}]%
        {Ludewig2018}
\bibfield{author}{\bibinfo{person}{Malte Ludewig} {and}
  \bibinfo{person}{Dietmar Jannach}.} \bibinfo{year}{2018}\natexlab{}.
\newblock \showarticletitle{Evaluation of Session-based Recommendation
  Algorithms}.
\newblock \bibinfo{journal}{\emph{User-Modeling and User-Adapted Interaction}}
  \bibinfo{volume}{28}, \bibinfo{number}{4--5} (\bibinfo{year}{2018}),
  \bibinfo{pages}{331--390}.
\newblock


\bibitem[\protect\citeauthoryear{Ludewig, Mauro, Latifi, and Jannach}{Ludewig
  et~al\mbox{.}}{2019}]%
        {LudewigMauro2019}
\bibfield{author}{\bibinfo{person}{Malte Ludewig}, \bibinfo{person}{Noemi
  Mauro}, \bibinfo{person}{Sara Latifi}, {and} \bibinfo{person}{Dietmar
  Jannach}.} \bibinfo{year}{2019}\natexlab{}.
\newblock \showarticletitle{Performance Comparison of Neural and Non-Neural
  Approaches to Session-based Recommendation}. In
  \bibinfo{booktitle}{\emph{Proceedings RecSys '19}}.
\newblock
\showISBNx{978-1-4503-6243-6/19/09}
\urldef\tempurl%
\url{https://doi.org/10.1145/3298689.3347041}
\showDOI{\tempurl}


\bibitem[\protect\citeauthoryear{Maksai, Garcin, and Faltings}{Maksai
  et~al\mbox{.}}{2015}]%
        {Maksai:2015:POP:2792838.2800184}
\bibfield{author}{\bibinfo{person}{Andrii Maksai}, \bibinfo{person}{Florent
  Garcin}, {and} \bibinfo{person}{Boi Faltings}.}
  \bibinfo{year}{2015}\natexlab{}.
\newblock \showarticletitle{Predicting Online Performance of News Recommender
  Systems Through Richer Evaluation Metrics}. In
  \bibinfo{booktitle}{\emph{Proceedings RecSys '15}}.
  \bibinfo{pages}{179--186}.
\newblock


\bibitem[\protect\citeauthoryear{Manotumruksa, Macdonald, and
  Ounis}{Manotumruksa et~al\mbox{.}}{2018}]%
        {manotumruksa2018contextualattention}
\bibfield{author}{\bibinfo{person}{Jarana Manotumruksa}, \bibinfo{person}{Craig
  Macdonald}, {and} \bibinfo{person}{Iadh Ounis}.}
  \bibinfo{year}{2018}\natexlab{}.
\newblock \showarticletitle{A Contextual Attention Recurrent Architecture for
  Context-Aware Venue Recommendation}. In \bibinfo{booktitle}{\emph{Proceedings
  SIGIR '18}}. \bibinfo{pages}{555--564}.
\newblock


\bibitem[\protect\citeauthoryear{Ning and Karypis}{Ning and Karypis}{2011}]%
        {ning2011SLIM}
\bibfield{author}{\bibinfo{person}{Xia Ning} {and} \bibinfo{person}{George
  Karypis}.} \bibinfo{year}{2011}\natexlab{}.
\newblock \showarticletitle{{SLIM: Sparse linear methods for top-n recommender
  systems}}. In \bibinfo{booktitle}{\emph{Proceedings ICDM '11}}.
  \bibinfo{pages}{497--506}.
\newblock


\bibitem[\protect\citeauthoryear{Paudel, Christoffel, Newell, and
  Bernstein}{Paudel et~al\mbox{.}}{2017}]%
        {paudel2017Rp3beta}
\bibfield{author}{\bibinfo{person}{Bibek Paudel}, \bibinfo{person}{Fabian
  Christoffel}, \bibinfo{person}{Chris Newell}, {and} \bibinfo{person}{Abraham
  Bernstein}.} \bibinfo{year}{2017}\natexlab{}.
\newblock \showarticletitle{Updatable, Accurate, Diverse, and Scalable
  Recommendations for Interactive Applications}.
\newblock \bibinfo{journal}{\emph{ACM Transactions on Interactive Intelligent
  Systems}} \bibinfo{volume}{7}, \bibinfo{number}{1} (\bibinfo{year}{2017}),
  \bibinfo{pages}{1}.
\newblock


\bibitem[\protect\citeauthoryear{Plesser}{Plesser}{2017}]%
        {Plesser2018}
\bibfield{author}{\bibinfo{person}{Hans~Ekkehard Plesser}.}
  \bibinfo{year}{2017}\natexlab{}.
\newblock \showarticletitle{Reproducibility vs. Replicability: A Brief History
  of a Confused Terminology}.
\newblock \bibinfo{journal}{\emph{Frontiers in Neuroinformatics}}
  \bibinfo{volume}{11}, \bibinfo{number}{76} (\bibinfo{year}{2017}).
\newblock


\bibitem[\protect\citeauthoryear{Quadrana, Cremonesi, and Jannach}{Quadrana
  et~al\mbox{.}}{2018}]%
        {QuadranaetalCSUR2018}
\bibfield{author}{\bibinfo{person}{Massimo Quadrana}, \bibinfo{person}{Paolo
  Cremonesi}, {and} \bibinfo{person}{Dietmar Jannach}.}
  \bibinfo{year}{2018}\natexlab{}.
\newblock \showarticletitle{Sequence-Aware Recommender Systems}.
\newblock \bibinfo{journal}{\emph{Comput. Surveys}} \bibinfo{volume}{51},
  \bibinfo{number}{4} (\bibinfo{year}{2018}), \bibinfo{pages}{1--36}.
\newblock


\bibitem[\protect\citeauthoryear{Rossetti, Stella, and Zanker}{Rossetti
  et~al\mbox{.}}{2016}]%
        {Rossetti:2016:COO:2959100.2959176}
\bibfield{author}{\bibinfo{person}{Marco Rossetti}, \bibinfo{person}{Fabio
  Stella}, {and} \bibinfo{person}{Markus Zanker}.}
  \bibinfo{year}{2016}\natexlab{}.
\newblock \showarticletitle{Contrasting Offline and Online Results when
  Evaluating Recommendation Algorithms}. In
  \bibinfo{booktitle}{\emph{Proceedings RecSys '16}}. \bibinfo{pages}{31--34}.
\newblock


\bibitem[\protect\citeauthoryear{Sachdeva, Gupta, and Pudi}{Sachdeva
  et~al\mbox{.}}{2018}]%
        {Sachdeva:2018:ANA:3240323.3240397}
\bibfield{author}{\bibinfo{person}{Noveen Sachdeva}, \bibinfo{person}{Kartik
  Gupta}, {and} \bibinfo{person}{Vikram Pudi}.}
  \bibinfo{year}{2018}\natexlab{}.
\newblock \showarticletitle{Attentive Neural Architecture Incorporating Song
  Features for Music Recommendation}. In \bibinfo{booktitle}{\emph{Proceedings
  RecSys '18}}. \bibinfo{pages}{417--421}.
\newblock


\bibitem[\protect\citeauthoryear{Said and Bellog\'{\i}n}{Said and
  Bellog\'{\i}n}{2014}]%
        {Said:2014:RTF:2645710.2645712}
\bibfield{author}{\bibinfo{person}{Alan Said} {and} \bibinfo{person}{Alejandro
  Bellog\'{\i}n}.} \bibinfo{year}{2014}\natexlab{}.
\newblock \showarticletitle{Rival: A Toolkit to Foster Reproducibility in
  Recommender System Evaluation}. In \bibinfo{booktitle}{\emph{Proceedings
  RecSys '14}}. \bibinfo{pages}{371--372}.
\newblock


\bibitem[\protect\citeauthoryear{Sarwar, Karypis, Konstan, and Riedl}{Sarwar
  et~al\mbox{.}}{2001}]%
        {sarwar2001item}
\bibfield{author}{\bibinfo{person}{Badrul Sarwar}, \bibinfo{person}{George
  Karypis}, \bibinfo{person}{Joseph Konstan}, {and} \bibinfo{person}{John
  Riedl}.} \bibinfo{year}{2001}\natexlab{}.
\newblock \showarticletitle{Item-based collaborative filtering recommendation
  algorithms}. In \bibinfo{booktitle}{\emph{Proceedings WWW '01}}.
  \bibinfo{pages}{285--295}.
\newblock


\bibitem[\protect\citeauthoryear{Sun, Yang, Zhang, Bozzon, Huang, and Xu}{Sun
  et~al\mbox{.}}{2018}]%
        {Sun:2018:RKG:3240323.3240361}
\bibfield{author}{\bibinfo{person}{Zhu Sun}, \bibinfo{person}{Jie Yang},
  \bibinfo{person}{Jie Zhang}, \bibinfo{person}{Alessandro Bozzon},
  \bibinfo{person}{Long-Kai Huang}, {and} \bibinfo{person}{Chi Xu}.}
  \bibinfo{year}{2018}\natexlab{}.
\newblock \showarticletitle{Recurrent Knowledge Graph Embedding for Effective
  Recommendation}. In \bibinfo{booktitle}{\emph{Proceedings RecSys '18}}.
  \bibinfo{pages}{297--305}.
\newblock


\bibitem[\protect\citeauthoryear{Tay, Anh~Tuan, and Hui}{Tay
  et~al\mbox{.}}{2018a}]%
        {tay2018latentrelationalmetric}
\bibfield{author}{\bibinfo{person}{Yi Tay}, \bibinfo{person}{Luu Anh~Tuan},
  {and} \bibinfo{person}{Siu~Cheung Hui}.} \bibinfo{year}{2018}\natexlab{a}.
\newblock \showarticletitle{Latent relational metric learning via memory-based
  attention for collaborative ranking}. In
  \bibinfo{booktitle}{\emph{Proceedings WWW '18}}. \bibinfo{pages}{729--739}.
\newblock


\bibitem[\protect\citeauthoryear{Tay, Tuan, and Hui}{Tay
  et~al\mbox{.}}{2018b}]%
        {tay2018multipointercoattention}
\bibfield{author}{\bibinfo{person}{Yi Tay}, \bibinfo{person}{Luu~Anh Tuan},
  {and} \bibinfo{person}{Siu~Cheung Hui}.} \bibinfo{year}{2018}\natexlab{b}.
\newblock \showarticletitle{Multi-Pointer Co-Attention Networks for
  Recommendation}. In \bibinfo{booktitle}{\emph{Proceedings SIGKDD '18}}.
  \bibinfo{pages}{2309--2318}.
\newblock


\bibitem[\protect\citeauthoryear{Tuan and Phuong}{Tuan and Phuong}{2017}]%
        {Tuan:2017:CNS:3109859.3109900}
\bibfield{author}{\bibinfo{person}{Trinh~Xuan Tuan} {and}
  \bibinfo{person}{Tu~Minh Phuong}.} \bibinfo{year}{2017}\natexlab{}.
\newblock \showarticletitle{3D Convolutional Networks for Session-based
  Recommendation with Content Features}. In
  \bibinfo{booktitle}{\emph{Proceedings RecSys '17}}.
  \bibinfo{pages}{138--146}.
\newblock


\bibitem[\protect\citeauthoryear{Vasile, Smirnova, and Conneau}{Vasile
  et~al\mbox{.}}{2016}]%
        {Vasile:2016:MPE:2959100.2959160}
\bibfield{author}{\bibinfo{person}{Flavian Vasile}, \bibinfo{person}{Elena
  Smirnova}, {and} \bibinfo{person}{Alexis Conneau}.}
  \bibinfo{year}{2016}\natexlab{}.
\newblock \showarticletitle{Meta-Prod2Vec: Product Embeddings Using
  Side-Information for Recommendation}. In
  \bibinfo{booktitle}{\emph{Proceedings RecSys '16}}.
  \bibinfo{pages}{225--232}.
\newblock


\bibitem[\protect\citeauthoryear{Wagstaff}{Wagstaff}{2012}]%
        {DBLP:journals/corr/abs-1206-4656}
\bibfield{author}{\bibinfo{person}{Kiri Wagstaff}.}
  \bibinfo{year}{2012}\natexlab{}.
\newblock \showarticletitle{Machine Learning that Matters}. In
  \bibinfo{booktitle}{\emph{Proceedings ICML '12}}. \bibinfo{pages}{529--536}.
\newblock


\bibitem[\protect\citeauthoryear{Wang and Blei}{Wang and Blei}{2011}]%
        {wang2011collaborativetopicmodeling}
\bibfield{author}{\bibinfo{person}{Chong Wang} {and} \bibinfo{person}{David~M
  Blei}.} \bibinfo{year}{2011}\natexlab{}.
\newblock \showarticletitle{Collaborative topic modeling for recommending
  scientific articles}. In \bibinfo{booktitle}{\emph{Proceedings KDD '11}}.
  \bibinfo{pages}{448--456}.
\newblock


\bibitem[\protect\citeauthoryear{Wang, Wang, and Yeung}{Wang
  et~al\mbox{.}}{2015}]%
        {wang2015collaborativedeeplearning}
\bibfield{author}{\bibinfo{person}{Hao Wang}, \bibinfo{person}{Naiyan Wang},
  {and} \bibinfo{person}{Dit-Yan Yeung}.} \bibinfo{year}{2015}\natexlab{}.
\newblock \showarticletitle{Collaborative deep learning for recommender
  systems}. In \bibinfo{booktitle}{\emph{Proceedings KDD '15}}.
  \bibinfo{pages}{1235--1244}.
\newblock


\bibitem[\protect\citeauthoryear{Wang, De~Vries, and Reinders}{Wang
  et~al\mbox{.}}{2006}]%
        {wang2006unifying}
\bibfield{author}{\bibinfo{person}{Jun Wang}, \bibinfo{person}{Arjen~P
  De~Vries}, {and} \bibinfo{person}{Marcel~JT Reinders}.}
  \bibinfo{year}{2006}\natexlab{}.
\newblock \showarticletitle{Unifying user-based and item-based collaborative
  filtering approaches by similarity fusion}. In
  \bibinfo{booktitle}{\emph{Proceedings SIGIR '06}}. \bibinfo{pages}{501--508}.
\newblock


\bibitem[\protect\citeauthoryear{Wang, Robertson, de~Vries, and Reinders}{Wang
  et~al\mbox{.}}{2008}]%
        {wang2008probabilistic}
\bibfield{author}{\bibinfo{person}{Jun Wang}, \bibinfo{person}{Stephen
  Robertson}, \bibinfo{person}{Arjen~P de Vries}, {and}
  \bibinfo{person}{Marcel~JT Reinders}.} \bibinfo{year}{2008}\natexlab{}.
\newblock \showarticletitle{Probabilistic relevance ranking for collaborative
  filtering}.
\newblock \bibinfo{journal}{\emph{Information Retrieval}} \bibinfo{volume}{11},
  \bibinfo{number}{6} (\bibinfo{year}{2008}), \bibinfo{pages}{477--497}.
\newblock


\bibitem[\protect\citeauthoryear{Wu, DuBois, Zheng, and Ester}{Wu
  et~al\mbox{.}}{2016}]%
        {wu2016collaborative}
\bibfield{author}{\bibinfo{person}{Yao Wu}, \bibinfo{person}{Christopher
  DuBois}, \bibinfo{person}{Alice~X Zheng}, {and} \bibinfo{person}{Martin
  Ester}.} \bibinfo{year}{2016}\natexlab{}.
\newblock \showarticletitle{Collaborative denoising auto-encoders for top-n
  recommender systems}. In \bibinfo{booktitle}{\emph{Proceedings WSDM '16}}.
  \bibinfo{pages}{153--162}.
\newblock


\bibitem[\protect\citeauthoryear{Xiao and Benbasat}{Xiao and Benbasat}{2007}]%
        {Xiao:2007:EPR:2017327.2017335}
\bibfield{author}{\bibinfo{person}{Bo Xiao} {and} \bibinfo{person}{Izak
  Benbasat}.} \bibinfo{year}{2007}\natexlab{}.
\newblock \showarticletitle{E-commerce Product Recommendation Agents: Use,
  Characteristics, and Impact}.
\newblock \bibinfo{journal}{\emph{MIS Quarterly}} \bibinfo{volume}{31},
  \bibinfo{number}{1} (\bibinfo{date}{March} \bibinfo{year}{2007}),
  \bibinfo{pages}{137--209}.
\newblock
\showISSN{0276-7783}


\bibitem[\protect\citeauthoryear{Zheng, Lu, Jiang, Zhang, and Yu}{Zheng
  et~al\mbox{.}}{2018}]%
        {Zheng:2018:SCF:3240323.3240343}
\bibfield{author}{\bibinfo{person}{Lei Zheng}, \bibinfo{person}{Chun-Ta Lu},
  \bibinfo{person}{Fei Jiang}, \bibinfo{person}{Jiawei Zhang}, {and}
  \bibinfo{person}{Philip~S. Yu}.} \bibinfo{year}{2018}\natexlab{}.
\newblock \showarticletitle{Spectral Collaborative Filtering}. In
  \bibinfo{booktitle}{\emph{Proceedings RecSys '18}}.
  \bibinfo{pages}{311--319}.
\newblock


\end{thebibliography}

\end{document}